\documentclass[twocolumn, reprint, superscriptaddress, amsmath,amssymb,aps,prl]{revtex4-2}
\usepackage{graphicx}
\usepackage{dcolumn}
\usepackage{bm}
\usepackage[hidelinks,colorlinks=true,linkcolor=blue,citecolor=blue]{hyperref}
\usepackage[mathlines]{lineno}
\usepackage{mathrsfs}   
\usepackage{calligra}   
\DeclareMathAlphabet{\mathcalligra}{T1}{calligra}{m}{n}

\usepackage{braket}
\usepackage{color}
\usepackage{float}
\usepackage{amsmath}
\usepackage[dvipsnames]{xcolor}

\begin{document}

\title{Non-Resonant Boundary Time Crystals from Quantum Synchronization Breakdown}

\affiliation{Shenzhen Key Laboratory of Ultraintense Laser and Advanced Material Technology, Center for Intense Laser Application Technology, and 
College of Engineering Physics, Shenzhen Technology University, Shenzhen 518118, China}
\affiliation{Computational Materials Science Research Team, RIKEN Center for Computational Science (R-CCS), Kobe, Hyogo 650-0047, Japan}
\affiliation{Department of Physics, Renmin University of China, Beijing, 100872, China}

\affiliation{Beijing National Laboratory for Condensed Matter Physics,
Institute of Physics, Chinese Academy of Sciences, Beijing 100190, China}
\affiliation{School of Physical Sciences, University of Chinese Academy of Sciences, Beijing 100049, China}

\author{Jun Wang}
\thanks{These authors contribute equally to this work.}
\affiliation{Shenzhen Key Laboratory of Ultraintense Laser and Advanced Material Technology, Center for Intense Laser Application Technology, and 
College of Engineering Physics, Shenzhen Technology University, Shenzhen 518118, China}

\author{Shu Yang}
\thanks{These authors contribute equally to this work.}
\affiliation{Shenzhen Key Laboratory of Ultraintense Laser and Advanced Material Technology, Center for Intense Laser Application Technology, and 
College of Engineering Physics, Shenzhen Technology University, Shenzhen 518118, China}

\author{Zeqing Wang}
\affiliation{Computational Materials Science Research Team, RIKEN Center for Computational Science (R-CCS), Kobe, Hyogo 650-0047, Japan}
\affiliation{Quantum Science Center of Guangdong-Hongkong-Macao Greater Bay Area (Guangdong), Shenzhen 518045, China}

\author{Ran Qi}
\affiliation{Department of Physics, Renmin University of China, Beijing, 100872, China}

\author{Haiping Hu}
\affiliation{Beijing National Laboratory for Condensed Matter Physics,
Institute of Physics, Chinese Academy of Sciences, Beijing 100190, China}
\affiliation{School of Physical Sciences, University of Chinese Academy of Sciences, Beijing 100049, China}

\author{Weidong Li}
\email{Corresponding author: liweidong@sztu.edu.cn}
\affiliation{Shenzhen Key Laboratory of Ultraintense Laser and Advanced Material Technology, Center for Intense Laser Application Technology, and 
College of Engineering Physics, Shenzhen Technology University, Shenzhen 518118, China}
 
\author{Jianwen Jie}
\email{Corresponding author: Jianwen.Jie1990@gmail.com}
\affiliation{Shenzhen Key Laboratory of Ultraintense Laser and Advanced Material Technology, Center for Intense Laser Application Technology, and 
College of Engineering Physics, Shenzhen Technology University, Shenzhen 518118, China}

\date{\today}

\begin{abstract}
Quantum synchronization (QS) in dissipative systems is often inferred from smooth phase locking, leaving open whether its breakdown constitutes a genuine nonequilibrium transition.
Here we introduce a Liouvillian framework that classifies driven–dissipative dynamics by the structure of the undriven dissipative background and show that QS breaks down via a Hopf-type dynamical phase transition into a boundary time crystal (BTC).
The character of this transition is determined by the background attractor: systems with a self-sustained oscillator (SSO) support robust non-resonant BTCs, whereas those with a polar fixed point (PFP) sustain BTCs only at resonance and lose them under detuning.
We identify sharp dynamical and spectral signatures of the QS–BTC transition and thereby establish, within U(1)-symmetric collective-spin Lindbladians driven by a single coherent tone, a background-based allowed/forbidden criterion that unifies QS, its breakdown, and time-crystalline order within a single Liouvillian framework.
\end{abstract}

\maketitle

{\em Introduction.---}A central question for quantum synchronization (QS) in dissipative quantum systems is how it breaks down: through a genuine dynamical transition, or via a crossover into an unsynchronized regime? QS---a distinctive manifestation of quantum coherence and correlations~\cite{galve2017quantum,PRA2013Spin}---has been realized on a wide range of platforms~\cite{PRL2013VdP,PRL2014VdP,PRL2019VdP,PRE2024Sudler,PRL2013OM1,PRL2018VdP,PRA2018coldatom,PRA2015QED,Weiss_2016,PRE2012OM,PRA2014OM,PRL2012NOexp,PRL2018Spin1,PRA2020two,PRL2018QN,Zhang2023PRR,PRA2022nuclear,PRR2020SPin1,Solanki2023PRA,PRA2024Tobia,PRB2009QED,PRB2024No_Go,Tan2022halfintegervs,PRR2020hybrid} with close ties to entanglement~\cite{he_entanglement_2024}, metrology~\cite{PRA2025QSandQFI,shen_fisher_2023}, topology~\cite{wachtler2023topological,wachtler2024topological,mr1f-v8cv}, and nonreciprocal dynamics~\cite{NC2025lai}.  However, existing studies typically diagnose QS from smooth changes in phase distributions, without a sharp spectral or dynamical signature of its breakdown \cite{PRL2018Spin1,PRA2020two,PRL2018QN,Zhang2023PRR,PRA2022nuclear,PRR2020SPin1}.  As a result, the boundary between QS and its breakdown is blurred, raising the question of whether QS constitutes a distinct nonequilibrium phenomenon.

In dissipative spin systems the Liouvillian can support qualitatively different background attractors, most notably polar fixed points (PFPs) and self-sustained oscillators (SSOs). In phase space, SSOs correspond to stable limit cycles. QS arises from phase-locking between such SSOs, induced by either external drives or interactions. The same limit-cycle structure also provides the dynamical backbone for time-crystalline phases~\cite{RMP2023Norman,PRL2012CTC,PRL2012QTC,Yang2025}. Boundary time crystal (BTC) provides a paradigmatic realization of continuous time-crystalline order in driven--dissipative spin ensembles~\cite{PRL2018BTC,Montenegro2023,PRB2021dbtc,PRB2022BTC,das2025stabilizingboundarytimecrystals,PRL2025BTC,PRB2021dbtc,PRB2021SCbtc,SP2025BTC}: collective dissipation stabilizes a PFP background, which undergoes a transition to persistent oscillations under resonant driving. While such BTCs are robust against heating instabilities~\cite{Review2020DTC}, they are intrinsically resonance-dependent and melt rapidly under detuning. This raises a structural question: which undriven dissipative backgrounds can support oscillatory phases that remain stable away from resonance, i.e., non-resonant BTC?

\begin{figure}[b]
    \centering
    \includegraphics[width=8.6cm]{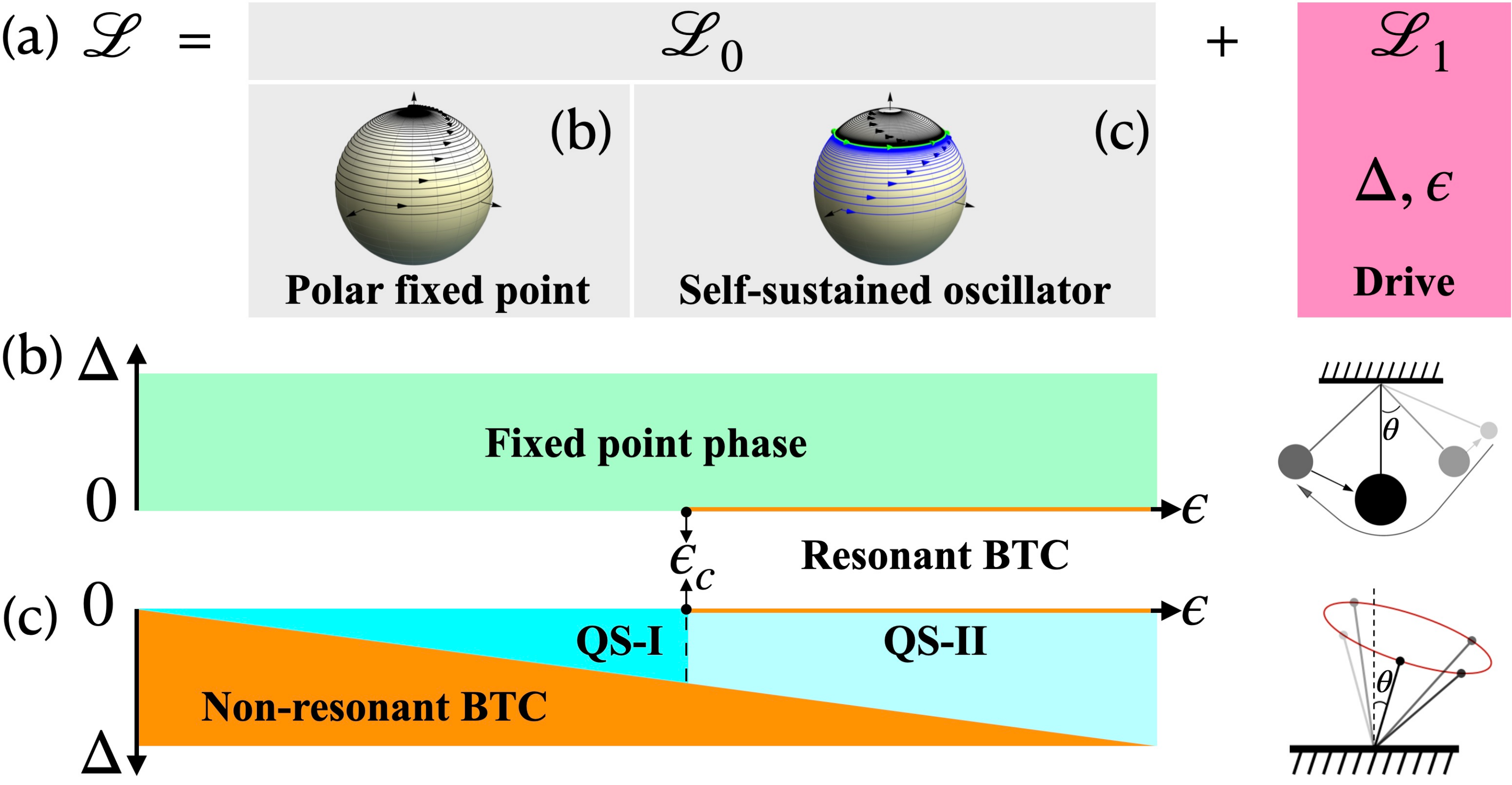}
    \caption{\label{fig1} (a) The Liouvillian of a driven--dissipative spin system is decomposed into 
$\mathcal{L}_0$, which governs the U(1)-symmetric undriven dissipative background, 
and $\mathcal{L}_1$, which represents the symmetry-breaking drive.
(b,c) Schematic phase diagrams in the rotating frame for cases where the background steady state set by $\mathcal{L}_0$ is a polar fixed point (b) or a self-sustained oscillator (c). 
For a PFP background, only a resonant BTC appears when drive strength $\epsilon>\epsilon_c$ and collapses under detuning $\Delta$, whereas an SSO background supports a non-resonant BTC over a finite detuning window. 
PFP behaves as a heavily damped pendulum, whereas SSO responds in a gyroscope-like manner, sustaining rotation.}
\end{figure}

A general framework identifying the role of the background attractor—SSO versus PFP—in connecting QS and BTC is still lacking. Here we develop such a Liouvillian framework for driven--dissipative spin systems and show that QS breaks down via a dynamical phase transition (DPT) into a BTC phase rather than a smooth crossover. Crucially, whether the resulting BTC survives detuning is dictated by the undriven background attractor: BTCs supported by PFP backgrounds melt under detuning from resonance [see Fig.~\ref{fig1}(b)], whereas those built on SSO backgrounds persist and realize robust non-resonant BTC phases [see Fig.~\ref{fig1}(c)]. These results reveal a common Liouvillian structure connecting QS and time-crystalline order, and delineate when non-resonant BTCs are fundamentally allowed or ruled out.

\emph{Main Results---}
QS breakdown is a Hopf-type DPT into a BTC within U(1)-symmetric collective-spin Lindbladians driven by a single coherent tone. 
In this setting, non-resonant BTCs emerge only in the presence of an undriven SSO background, whereas backgrounds supporting only PFP attractors fail to sustain them under detuning.

As illustrated in Fig.~\ref{fig1}(a), we decompose the Lindblad master equation as
\begin{equation}
\frac{d\hat{\rho}}{dt}=\mathscr{L}[\hat{\rho}]=\mathscr{L}_{0}[\hat{\rho}]+\mathscr{L}_{1}[\hat{\rho}].
\label{eq:generalME}
\end{equation}
The undriven part reads $\mathscr{L}_0[\hat{\rho}]=-i[H_0,\hat{\rho}]+\mathscr{D}[\hat{\rho}]$ with
$H_0=\omega_0 \hat{S}^z$ for a collective-spin $S=N/2$ built from $N$ identical spin-$1/2$ particles,
$\hat{S}^{\alpha=x,y,z}=\sum_{i=1}^N \hat{\sigma}_i^{(\alpha)}/2$,
and
$\mathscr{D}[\hat{\rho}]=\sum_{k=1}^{M}\!\left(\hat{L}_k \hat{\rho}\hat{L}_k^\dagger-\{\hat{L}_k^\dagger \hat{L}_k,\hat{\rho}\}/2\right)$,
where the jump operators $\hat{L}_k$ conserve the $U(1)$ symmetry of $H_0$.

In the Schr\"odinger picture, $\mathscr{L}_0$ selects two distinct classes of background attractors:
(i) an always-present PFP, given by the stationary extremal state $|S,\pm S\rangle$ satisfying
$\mathscr{L}_0[|S,\pm S\rangle\langle S,\pm S|]=0$; here no spontaneously selected azimuthal phase exists, so phase locking (and thus QS) is not applicable.
(ii) for suitable dissipation, $\mathscr{L}_0$ may additionally support a SSO, i.e., a dissipation-stabilized precessional limit cycle away from full polarization~\cite{SSO_footnote}. This oscillatory attractor carries a free azimuthal phase $\phi$ underlying QS, and is spectrally diagnosed in the thermodynamic limit by purely imaginary Liouvillian eigenvalues.

Nontrivial behavior arises from the interplay between $\mathscr{L}_0$ and a noncommuting coherent drive
$\mathscr{L}_1[\hat{\rho}]=-i[\hat{H}_d(t),\hat{\rho}]$, e.g., $\hat{H}_d(t)=-\epsilon\cos(\omega_d t)\hat{S}^x$.
The drive explicitly breaks the $U(1)$ symmetry of $\mathscr{L}_0$, enabling phase locking of the SSO to the drive; the loss of phase locking indicates QS breakdown.

Our main results are summarized in Fig.~\ref{fig1}(b,c), which show schematic rotating-frame phase diagrams for PFP and SSO backgrounds, respectively.
PFP backgrounds behave as heavily damped pendula, suppressing sustained phase rotation under detuning, whereas SSO backgrounds exhibit a gyroscope-like response that enables persistent rotation.
Along the resonance line ($\Delta=0$), increasing the drive strength $\epsilon$ across a critical value $\epsilon_c$ yields a Hopf-type DPT~\cite{PRL2018BTC,Montenegro2023}.
For an SSO background, this transition marks QS breakdown, taking the system from a phase-locked synchronized regime (QS-I) into a phase-non-locked resonant BTC; for a PFP background it creates a BTC out of an otherwise trivial stationary state.
In the thermodynamic limit, this loss of phase locking coincides with the onset of persistent oscillations, establishing BTC order.
Away from resonance ($\Delta\neq0$), the two backgrounds differ qualitatively: for PFPs, the resonant BTC rapidly melts under detuning into a trivial fixed-point state, whereas SSOs support a non-resonant BTC over a detuning window already for $\epsilon>0$.
At fixed finite detuning, increasing $\epsilon$ further induces a DPT from the non-resonant BTC to a synchronized phase (QS-I) with monotonic relaxation, followed at stronger drive by a crossover to a second synchronized regime (QS-II) exhibiting oscillatory relaxation.

\begin{figure}[t]
    \centering
    \includegraphics[width=8.63cm]{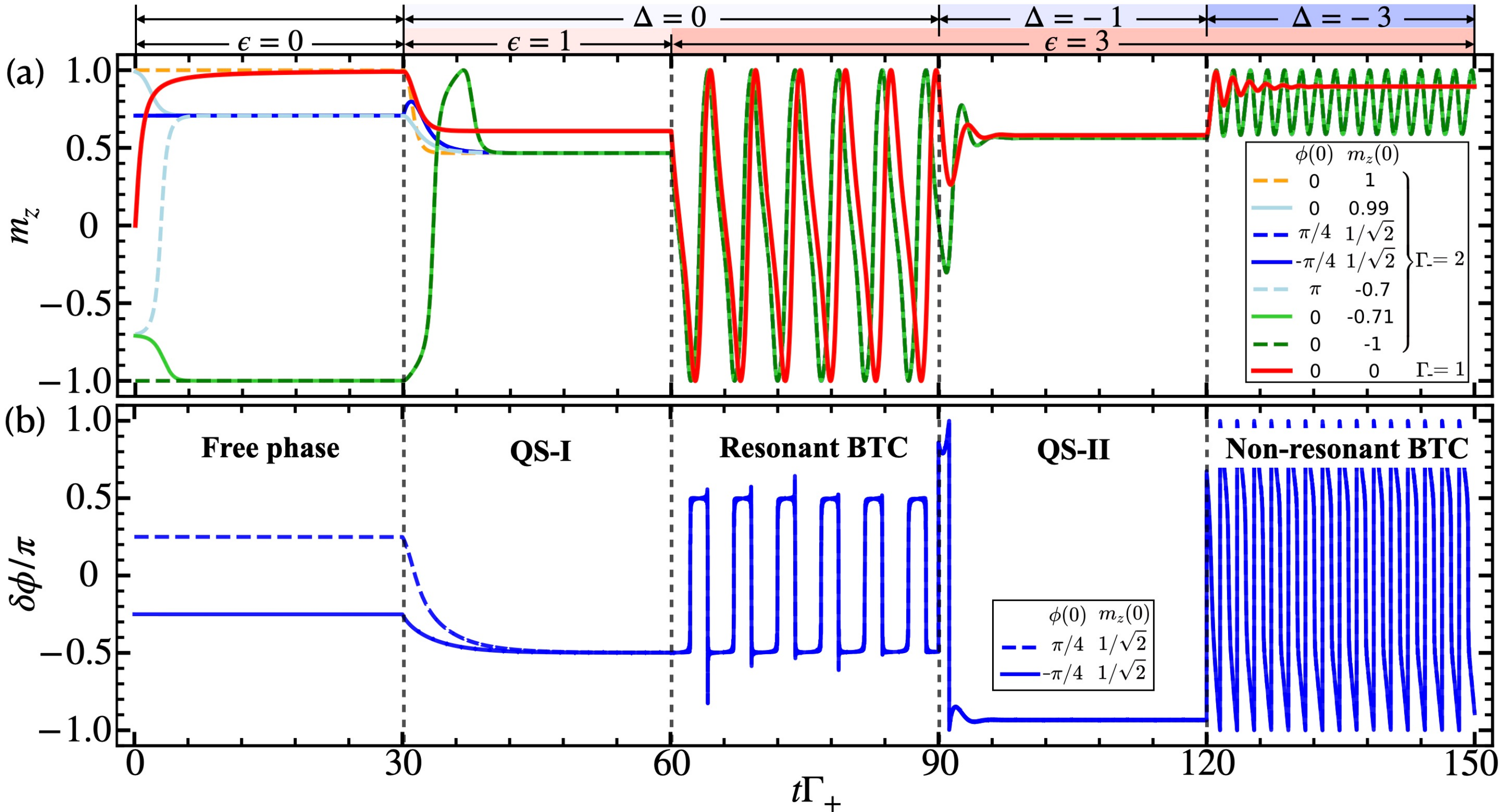}
\caption{\label{fig2}Dynamics of the minimal collective-spin model under the driving protocol shown on top: (a) magnetization \(m_z(t)\), comparing a PFP-only background (\(\Gamma_-=1\)) with an SSO-hosting background (\(\Gamma_-=2\)); (b) relative phase \(\delta\phi(t)=\phi(t)-[\omega_d t]_{2\pi}\) for two initial conditions on the SSO manifold, diagnosing phase locking and its breakdown. Parameters: \(\omega_0=20\pi\), \(\Gamma_+=1\).}
\end{figure}

These findings establish that the presence of an SSO background constitutes the structural prerequisite—and thus a sharp allowed/forbidden criterion—for non-resonant BTCs within our framework. To make this criterion explicit microscopically, we illustrate the background-attractor classification in a minimal collective-spin model realizing both PFP and SSO attractors within a unified setting.

\emph{Minimal Collective-Spin Illustration.---}The model features two collective dissipation channels \cite{PRL2025Model}: a linear gain
\( \hat{L}_+ = \sqrt{\Gamma_{+}/S}\,\hat{S}_+ \)
and a nonlinear decay
\( \hat{L}_- = \sqrt{\Gamma_{-}/S^3}\,\hat{S}_- \hat{S}_z \),
with rates  \( \Gamma_{\pm} \) and
\( \hat{S}_\pm=\hat{S}_x \pm i\hat{S}_y \).
The \(S\)-dependent normalization of \( \hat{L}_\pm \) ensures a well-defined thermodynamic limit as \(N\to\infty\).
In this limit, applying a mean-field factorization together with the rotating-wave approximation to Eq.~(\ref{eq:generalME}),
we obtain the nonlinear equations for
\( m_\alpha \equiv \langle \hat{S}^\alpha\rangle/S \)
in the frame rotating at the drive frequency \( \omega_d \):
\begin{align}\label{neom}
 \frac{d m_x}{dt} &= -\Delta m_y - (\Gamma_{+} - \Gamma_{-} m_z^2)\, m_x m_z, \nonumber \\
 \frac{d m_y}{dt} &= \Delta m_x +\frac{\epsilon}{2} m_z - (\Gamma_{+} - \Gamma_{-} m_z^2)\, m_y m_z,  \\
 \frac{d m_z}{dt} &= -\frac{\epsilon}{2} m_y + (\Gamma_{+} - \Gamma_{-} m_z^2)\,(1 - m_z^2), \nonumber
\end{align}
where \( \Delta=\omega_0-\omega_d \).
The dynamics is confined to the Bloch sphere, \( \sum_{\alpha=x,y,z} m_\alpha^2 = 1 \), provided the initial state satisfies this constraint~\cite{footnote_SM}.

As shown in Fig.~\ref{fig2}(a), the undriven attractor structure is controlled by the damping ratio \( \Gamma_{+}/\Gamma_{-} \).
For \( \Gamma_{+}\ge \Gamma_{-} \), the system relaxes to the PFP at the north pole, \(m_z=+1\) (thick red line).
For \( \Gamma_{+}<\Gamma_{-} \), this north-pole PFP becomes unstable and an additional SSO emerges at
\( m_z^{\mathrm{SSO}}=\sqrt{\Gamma_{+}/\Gamma_{-}} \) (blue lines), while trajectories with sufficiently large initial deviations (\( m_z<-m_z^{\mathrm{SSO}} \)) relax to the south-pole PFP (green lines). We thus compare driven dynamics for two representative backgrounds: a PFP-only case (\( \Gamma_{-}/\Gamma_{+}=1 \)) and an SSO-hosting case (\( \Gamma_{-}/\Gamma_{+}=2 \)).

With the drive on, the long-time dynamics exhibits distinct attractors across the \((\epsilon,\Delta)\) plane.
On resonance (\(\Delta=0\)), increasing \(\epsilon\) induces a transition from a stationary fixed point to a stable limit cycle at a frequency much lower than the drive, realizing a resonant BTC~\cite{PRL2018BTC,Montenegro2023}.
Away from resonance (\(\Delta\neq0\)), this BTC is fragile on a PFP-only background: oscillations are damped out and the dynamics returns to a trivial fixed point.
By contrast, on an SSO background persistent oscillations re-emerge at sufficiently large detuning, yielding a non-resonant BTC.

QS is meaningfully defined only on an SSO background, where the azimuthal phase \(\phi(t)=\arg(m_x+i m_y)\) is well defined. Accordingly, Fig.~\ref{fig2}(b) shows the relative phase \(\delta\phi(t)=\phi(t)-[\omega_d t]_{2\pi}\) for two initial conditions on the SSO manifold with \(m_z(0)=m_z^{\mathrm{SSO}}\) and \(\phi(0)=\pm\pi/4\).
Without driving, this limit-cycle attractor retains its free azimuthal phase.
Under weak resonant driving, \(\delta\phi(t)\) relaxes exponentially to a constant (QS-I).
For stronger resonant driving, phase locking breaks down and \(\delta\phi(t)\) exhibits oscillatory, signaling a resonant BTC.
With detuning, two behaviors emerge: for moderate \(|\Delta|\), \(\delta\phi(t)\) approaches a locked phase via damped oscillations (QS-II), whereas for sufficiently large \(|\Delta|\) phase locking is destabilized and \(\delta\phi(t)\) becomes persistently oscillatory, identifying a non-resonant BTC.

\begin{figure}[t]
    \centering
    \includegraphics[width=8.6cm]{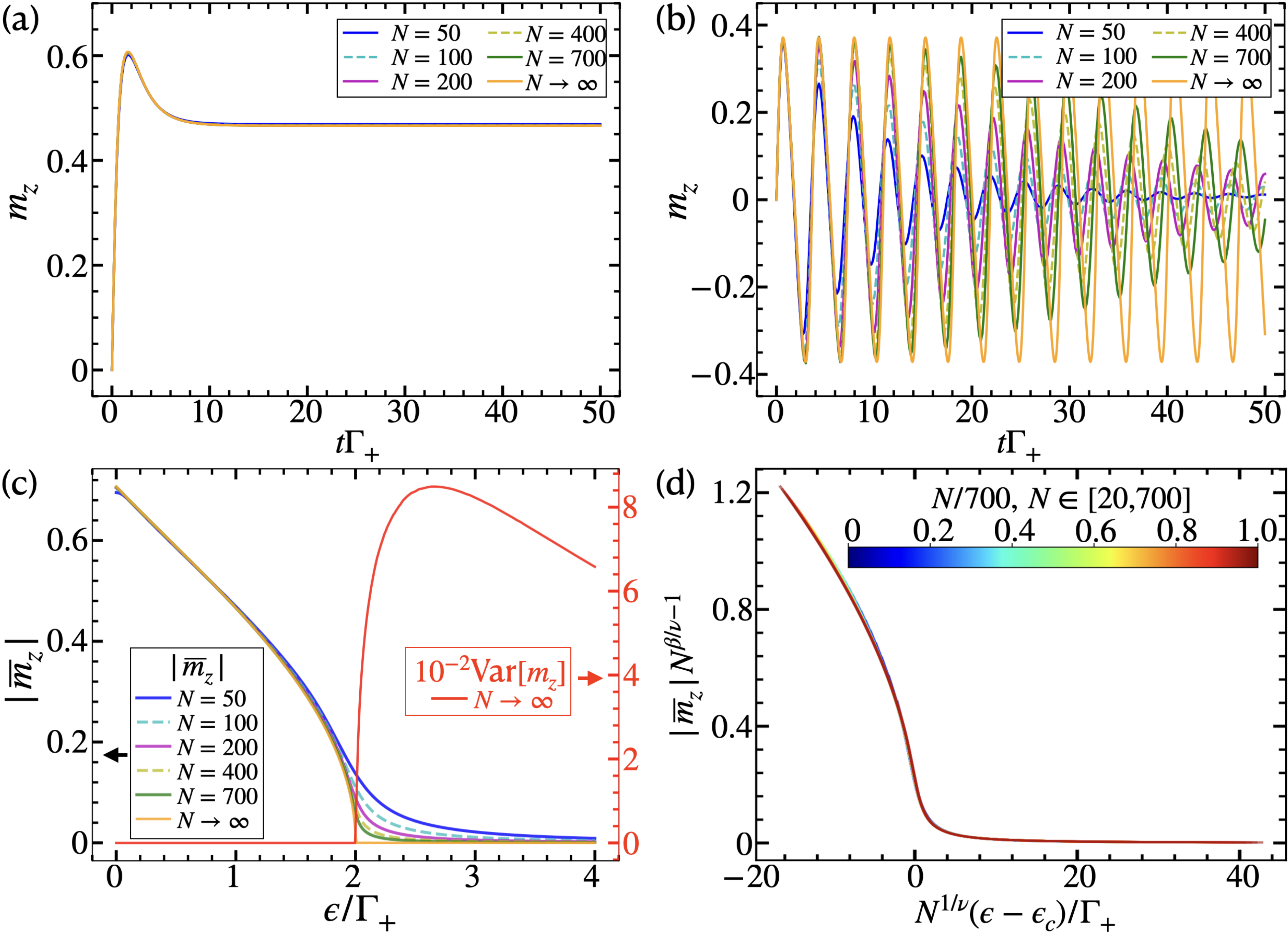}
        \caption{\label{fig3} Verification of the QS-BTC dynamical phase transition under resonant driving. (a–b) Time evolution of the magnetization \( m_z(t) \) for various system sizes \( N \), initialized at \( (m_x, m_y, m_z) = (1, 0, 0) \), under (a) weak driving \( \epsilon = 1 \) and (b) strong driving \( \epsilon = 4 \).  
(c) The nonanalytic behavior of the dynamical order parameters—the time-averaged magnetization \( \overline{m}_z \) and its variance \( \text{Var}[m_z] \)—as the system size increases, indicates a DPT from QS to BTC.  
(d) Universal data collapse in the finite-size scaling of \( \overline{m}_z \) confirms a DPT. Other parameters: \(\omega_0=\omega_d=20\pi,~\Gamma_{+} = 1,~\Gamma_{-} = 2 \).}
\end{figure}

{\em Dynamical QS--BTC Transition---}
Figure~\ref{fig3}(a--b) shows \(m_z(t)\) on both sides of the transition under resonant driving,
comparing finite-size results (\(N\in[20,800]\)) obtained by solving the Lindblad master equation
[Eq.~(\ref{eq:generalME})]~\cite{pythonpackage}
with the mean-field (thermodynamic-limit) dynamics described by the nonlinear equations [Eq.~(\ref{neom})].
In the QS phase (\(\epsilon<\epsilon_c\)), \(m_z(t)\) relaxes to a stationary value \(m_{z,\mathrm{ss}}\).
In contrast, in the BTC phase (\(\epsilon>\epsilon_c\)), \(m_z(t)\) exhibits oscillations whose damping weakens with increasing \(N\),
consistent with a stable limit cycle in the thermodynamic limit. The phase-locking diagnostic $\delta\phi(t)$ yields a critical point consistent with that obtained from the order-parameter scaling, indicating that the breakdown of QS coincides with the onset of the DPT.

These contrasting long-time behaviors motivate the time-averaged magnetization,
\(\overline{m}_z=\Delta t^{-1}\int_{t_0}^{t_0+\Delta t} m_z(t)\,dt\),
and its fluctuations,
\(\mathrm{Var}[m_z]=\Delta t^{-1}\int_{t_0}^{t_0+\Delta t} m_z^2(t)\,dt-\overline{m}_z^{\,2}\),
as dynamical order parameters for the QS--BTC transition.
We choose \(t_0\gg\tau\) after transients (with \(\tau\) the relaxation time) and \(\Delta t\gg 1/f\) to average over many periods of oscillation frequency \(f\).
As shown in Fig.~\ref{fig3}(c), \(|\overline{m}_z|\) develops an increasingly sharp feature near \(\epsilon_c\) with growing \(N\),
consistent with a nonanalytic behavior in the thermodynamic limit.
This is further supported by the pronounced feature in \(\mathrm{Var}[m_z]\), providing evidence for a DPT~\cite{Montenegro2023}.

To further corroborate a well-defined critical point, we perform a finite-size scaling analysis~\cite{pythonpackage,Montenegro2023}.
Using a finite-size scaling ansatz for the dynamical order parameter,
\(\overline{m}_z = N^{-\beta/\nu} f\!\left[N^{1/\nu}(\epsilon-\epsilon_c)\right]\),
we obtain an excellent data collapse for system sizes \(N\in[20,800]\), as shown in Fig.~\ref{fig3}(d). The collapse yields the critical point \( \epsilon_c = 1.98 \pm 0.021 \), along with critical exponents \( \nu = 1.74 \pm 0.079 \) and \( \beta = 0.47 \pm 0.11 \).
The extracted \(\epsilon_c\) is consistent with the linear-stability analysis reported in Ref.~\cite{footnote_SM}, which gives \(\epsilon_c = 2.0\).
Together, these results support a continuous (second-order) DPT.

Having established a well-defined critical point on resonance in Fig.~\ref{fig3}, we now extend the same dynamical diagnostics to the \((\epsilon,\Delta)\) plane by mapping out an Arnold-tongue-like phase diagram for the SSO background.

\begin{figure}[t]
    \centering
    \includegraphics[width=8.63cm]{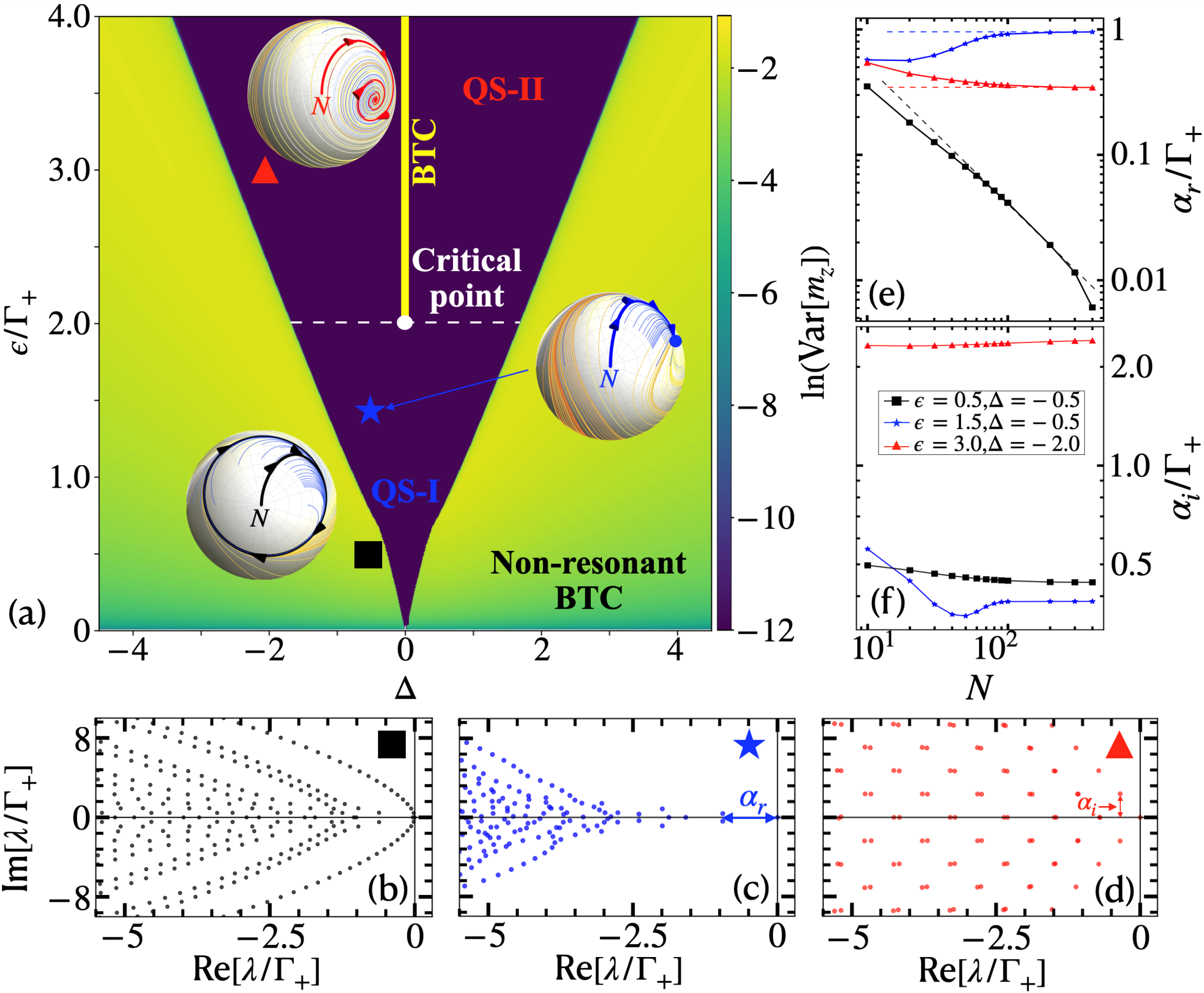}
    \caption{\label{fig4}  (a) Arnold tongue phase diagram showing the boundary between two phase-locked Quantum synchronization (QS) regimes and BTC regimes.
    (b) Bloch trajectories and Liouvillian spectra for the three representative points marked in Fig.~\ref{fig3}(a): square (BTC), star (QS-I), and triangle (QS-II), with \( N = 200 \). 
    Each thick trajectory is initialized at the north pole to highlight their distinct dynamical behaviors: persistent oscillations in BTC, exponential relaxation in QS-I, and oscillatory decay in QS-II.
    (d-e) Finite-size scaling of the Liouvillian gap $\alpha=\alpha_r \pm i\alpha_i$ for the three cases in (b). Dashed lines are guides to the eye. 
    Other parameters: \( \omega_0=20\pi,~ \Gamma_+ = 1 \), \( \Gamma_- = 2 \).}
\end{figure}

\emph{Arnold Tongue and Liouvillian Signatures---}
As shown in Fig.~\ref{fig4}(a), the phase boundary is obtained from the long-time dynamics using the variance \(\mathrm{Var}[m_z]\) as a dynamical order parameter.
To connect this phase diagram to microscopic relaxation, we select three representative points in Fig.~\ref{fig4}(a) and compare their Bloch-sphere trajectories and Liouvillian spectra in Figs.~\ref{fig4}(b--d).
At the square point, the Bloch trajectory reveals bistability between a stable limit cycle and an attracting fixed point,
identifying the non-resonant BTC regime; correspondingly, the spectrum shows coexisting wedge- and parabolic-like structures.
In contrast, both the star and triangle points display a single attracting fixed point (QS), but with distinct relaxation:
the star point shows fast exponential decay with a predominantly wedge-shaped spectrum (QS-I),
whereas the triangle point exhibits slow oscillatory (spiral) decay with pronounced parabolic components superimposed on the wedge background (QS-II).
These two QS regimes are separated by a crossover in the relaxation spectrum.

As shown in Fig.~\ref{fig4}(e,f), the relaxation dynamics is captured by the finite-size scaling of the Liouvillian gap
$\alpha=\alpha_r \pm i\alpha_i$, associated with the leading decay mode. At the transition, $\alpha_r$ vanishes with system size while $\alpha_i$ remains finite, constituting a Hopf-type Liouvillian instability.
In the BTC regime, $\alpha_r$ decreases toward zero with increasing $N$ while $\alpha_i$ approaches a finite value (here $\alpha_i\simeq0.44\ll\omega_0$), consistent with persistent low-frequency oscillations in the thermodynamic limit.
By contrast, in both QS-I and QS-II, $\alpha_r$ approaches a finite value as $N$ increases, implying a gapped Liouvillian and relaxation to a stationary fixed point. Although the two QS regimes can display similar $\alpha_r$ at small $N$, their large-$N$ relaxation differs: QS-I exhibits predominantly exponential decay, whereas QS-II shows pronounced oscillatory (spiral) decay.

To quantify this distinction, we introduce the dimensionless ratio
$R\equiv|\alpha_i/\alpha_r|$.
QS-I corresponds to $R\ll1$, whereas QS-II is characterized by $R\gg1$.
In the BTC phase, $\alpha_r\to0$ in the thermodynamic limit, implying $R\to\infty$.
For $N=400$, we find representative values $R\simeq0.4$ (QS-I), $R\simeq7$ (QS-II), and $R\simeq73$ (BTC).
We interpret the QS-I/QS-II crossover as correlating with a redistribution of spectral weight between wedge-shaped and parabolic components,
reflected in the behavior of $\alpha_i$ as the oscillation scale of the decaying dynamics.

{\em Physics Picture---}
A unified understanding follows by rewriting Eqs.~(\ref{neom}) in spherical coordinates~\cite{footnote_SM}.
With \(\theta\in[0,\pi]\) the polar (latitude) angle and \(\phi\in[0,2\pi)\) the azimuthal phase, one obtains
\(\dot{\theta}=(\epsilon\sin\phi)/2-\sin\theta\bigl(\Gamma_+-\Gamma_-\cos^2\theta\bigr)\) and
\(\dot{\phi}=\Delta+(\epsilon\cot\theta\,\cos\phi)/2\).
Dissipation generates a nonlinear latitudinal restoring force, yielding either
PFPs at the north/south poles or a finite-tilt SSO.

The resonant drive activates azimuthal dynamics through a phase-selecting torque with
latitude amplification \(\cot\theta\), which diverges near PFPs but remains bounded on an SSO.
Consequently, weak driving tends to pin \(\phi\) near \(\pi/2\), maximizing the
drive-induced latitudinal thrust and balancing the restoring force.
At strong drive this balance breaks down: \(\phi\) undergoes phase self-trapping
(bounded oscillations around a biased phase) in an asymmetric effective potential induced by \(\cot\theta\).
The resulting rapidly oscillating but biased thrust acquires a finite time average,
stabilizing a limit cycle and producing a resonant BTC for both backgrounds considered.

Detuning adds a constant precession torque that destroys phase self-trapping:
over long timescales \(\phi\) explores \([0,2\pi)\), washing out \(\phi\)–\(\theta\)
correlations and suppressing the effective thrust.
As a result, the resonant BTC melts to a stationary fixed point or a phase-locked QS state.
On a PFP background, the divergent amplification near the stable pole renders the dynamics akin to a heavily damped pendulum,
such that the phase-locked fixed point remains robust over a wide detuning range.
By contrast, on an SSO background the bounded phase-selecting torque yields a
spinning-gyroscope--like response: sufficiently strong detuning destabilizes
the locked phase, induces sustained azimuthal oscillations, and establishes a
new self-consistent dynamical balance in \(\theta\), giving rise to a non-resonant BTC. This physical picture is consistent with the structural distinction between PFP and SSO backgrounds discussed above.

\emph{Conclusion.—}We have shown that QS breakdown in driven-dissipative collective-spin systems constitutes a genuine Hopf-type dynamical phase transition in the thermodynamic limit, rather than a smooth crossover. By classifying Liouvillian dynamics according to the structure of their undriven background attractors, we identify a structural principle: non-resonant BTCs emerge only from self-sustained oscillator (SSO) backgrounds and are generically excluded for polar fixed point (PFP) backgrounds. Persistent collective oscillations thus reflect a property of the underlying Liouvillian background structure rather than a model-dependent feature.

Crucially, the SSO background relies on amplitude stabilization provided by nonlinear dissipation and is not restricted to the thermodynamic limit. In finite systems—including finite-spin implementations \cite{PRL2018Spin1,PRL2018QN,PRA2020two,Zhang2023PRR,PRA2022nuclear}, quantum van der Pol oscillators~\cite{PRA2025QSBandNR,PRL2024Yongchunliu,PRA2024Doerte,li2026general}, and Jaynes-Cummings platforms \cite{PRA1992JCLaser,PRL2010JCLaser,xiao2026squeeze}—SSOs manifest as amplitude-stabilized, phase-neutral ring steady states. Although such systems do not exhibit many-body phase transitions in the strict sense, they share the same background structure identified here, providing a unified language that connects finite-system SSO dynamics with thermodynamic-limit BTC transitions. Engineered nonlinear dissipation providing amplitude stabilization has been experimentally realized in cold-atom spin-1 systems via polarization-controlled optical pumping~\cite{PRL2020exp} and in trapped-ion realizations of quantum van der Pol oscillators through two-phonon damping induced by second-order red-sideband driving with rapid spin reset~\cite{QVDP_2025_Lin,liu2025}. Related nonlinear loss mechanisms have also been shown to stabilize otherwise unstable driven–dissipative light–matter systems, enabling well-defined steady-state phases in cavity QED platforms~\cite{PRL2025Sha}.

The same organizing principle extends naturally to networks of interacting SSOs~\cite{nadolny_nonreciprocal_2025,PRL2022seeding,46my-41ym,PRL2023MQSE,PRL2014atoms}. Within this framework, anisotropic couplings act as effective coherent drives~\cite{Luo2025,PRL2023Arrays,Li2024,Miller2024,lv2025polar}, offering a platform for collective nonequilibrium phenomena such as nonreciprocal active-quantum dynamics~\cite{nadolny_nonreciprocal_2025}, seeded time crystals~\cite{PRL2022seeding}, macroscopic oscillatory states~\cite{46my-41ym,PRL2023MQSE}, and chimera-like states~\cite{solanki_exotic_2024,PRE2015chimera,PRL2025chimera}. More broadly, the background–attractor viewpoint suggests that more complex Liouvillian attractors—such as quasiperiodic~\cite{PRL2025DTC,PRXQuantum2024DTC,Yusipov2019} or chaotic attractors~\cite{Dahan2022,PRR2025chaso,b6gq-z5nf}—may organize nonequilibrium behavior beyond PFPs and SSOs. Together, these results identify background-attractor structure as the organizing principle governing collective nonequilibrium oscillations.

\begin{acknowledgments}
We thank Peng Zhang, Zhigang Wu, Tao Shi, Xiao Yuan, Yingdan Wang and Stefano Chesi for valuable discussions. This work was supported by the National Natural Science Foundation of China (Grant No. 12575026, No. 12505027, No. 12474496 and No. 12547107), the Shenzhen Science and Technology Program (Grant No. JCYJ20250604145221028), the National Key Research and Development Program of China (Grant No. 2022YFA1405301, No. 2023YFA1406704 and No. 2022YFA1405800), the Natural Science Foundation of Top Talent of SZTU (Grant No. GDRC202202 and GDRC202312), and the Guangdong Provincial Quantum Science Strategic Initiative (Grants No. GDZX2305006 and No. GDZX2505001). 
\end{acknowledgments}

\bibliography{QSBTC_Refs.bib}

\global\long\def\id{\mathbbm{1}}
\global\long\def\ui{\mathbbm{i}}
\global\long\def\ud{\mathrm{d}}

\setcounter{equation}{0} \setcounter{figure}{0}
\setcounter{table}{0} 
\renewcommand{\theparagraph}{\bf}
\renewcommand{\thefigure}{S\arabic{figure}}
\renewcommand{\theequation}{S\arabic{equation}}

\onecolumngrid
\flushbottom
\newpage
\maketitle
\title{Supplementary Material:\\Non-Resonant Boundary Time Crystals from Quantum Synchronization Breakdown}
In this Supplemental Material we provide technical derivations and extended analyses supporting the main results of the manuscript.

First, we derive the mean-field equations starting from the microscopic Lindblad master equation in the Heisenberg picture and explicitly justify the large-$S$ factorization leading to Eq.~(2) of the main text. 
Second, we present a resonant linear-stability analysis of the mean-field dynamics and obtain the analytical threshold $\epsilon_c=2.0$ under the parameters used in Fig.~3. Finally, we derive the angular equations on the Bloch sphere and develop a unified physical picture in terms of latitude–phase dynamics, clarifying the pendulum-versus-gyroscope distinction between PFP and SSO backgrounds.

\section{System}\label{sec:2}

\subsection{Derivation of the mean-field equations}\label{sec:2_mf}
We start from the Lindblad master equation (in the rotating frame and within the RWA used in the main text),
\begin{equation}
\dot{\hat{\rho}}=-i[\hat H,\hat{\rho}]+\sum_{k=\pm}\left(\hat L_k\hat{\rho} \hat L_k^\dagger-\frac{1}{2}\{\hat L_k^\dagger \hat L_k,\hat{\rho}\}\right),
\end{equation}
whose Heisenberg-picture form for an arbitrary operator $\hat X$ reads
\begin{equation}
\frac{d\hat X}{dt}=i[\hat H,\hat X]+\frac{1}{2}\sum_{k=\pm}\Big(\hat L_k^\dagger[\hat X,\hat L_k]+[\hat L_k^\dagger,\hat X]\hat L_k\Big).
\label{eq:heis_lindblad}
\end{equation}
Here
\begin{equation}
\hat H=\Delta \hat S_z-\frac{\epsilon}{2}\hat S_x,\qquad \Delta=\omega_0-\omega_d,
\end{equation}
and the two collective jump operators are
\begin{equation}
\hat L_+=\sqrt{\frac{\Gamma_+}{S}}\hat S_+,\qquad
\hat L_-=\sqrt{\frac{\Gamma_-}{S^3}}\hat S_-\hat S_z ,
\label{eq:jumps}
\end{equation}
with $S=N/2$.
Throughout we use the $\mathfrak{su}(2)$ commutation relations
$[\hat S_i,\hat S_j]=i\epsilon_{ijk}\hat S_k$,
equivalently
\begin{equation}
[\hat S_z,\hat S_\pm]=\pm \hat S_\pm,\qquad [\hat S_+,\hat S_-]=2\hat S_z,
\qquad \hat S_x=\frac{\hat S_++\hat S_-}{2},\ \hat S_y=\frac{\hat S_+-\hat S_-}{2i}.
\label{eq:su2}
\end{equation}

\paragraph{Step 1: equations for $\hat S_+$ and $\hat S_z$.}
We first evaluate Eq.~\eqref{eq:heis_lindblad} for $\hat X=\hat S_+$.
The Hamiltonian part is straightforward:
\begin{align}
i[\hat H,\hat S_+]
&=i\Delta[\hat S_z,\hat S_+]-i\frac{\epsilon}{2}[\hat S_x,\hat S_+]
\nonumber\\
&=i\Delta \hat S_+ - i\frac{\epsilon}{2}\cdot \frac{1}{2}[\hat S_- ,\hat S_+]
= i\Delta \hat S_+ +\frac{i\epsilon}{2}\hat S_z .
\label{eq:Splus_H}
\end{align}
For the linear-gain channel $\hat L_+$ we obtain
\begin{align}
\frac{1}{2}\Big(\hat L_+^\dagger[\hat S_+,\hat L_+]+[\hat L_+^\dagger,\hat S_+]\hat L_+\Big)
&=\frac{1}{2}\frac{\Gamma_+}{S}\Big(\hat S_-[\hat S_+,\hat S_+]+[\hat S_-,\hat S_+]\hat S_+\Big)
\nonumber\\
&=\frac{1}{2}\frac{\Gamma_+}{S}\Big(0-2\hat S_z\hat S_+\Big)
= -\frac{\Gamma_+}{S}\hat S_z\hat S_+.
\label{eq:Splus_Lplus}
\end{align}

For the nonlinear-loss channel $\hat L_-=\sqrt{\Gamma_-/S^3}\,\hat S_-\hat S_z$ (so that $\hat L_-^\dagger=\sqrt{\Gamma_-/S^3}\,\hat S_z\hat S_+$),
a direct but lengthy commutator algebra yields an exact operator identity of the form
\begin{equation}
\frac{1}{2}\Big(\hat L_-^\dagger[\hat S_+,\hat L_-]+[\hat L_-^\dagger,\hat S_+]\hat L_-\Big)
=\frac{\Gamma_-}{2S^3}\Big(
2\hat S_z\hat S_+\hat S_z^2
+2\hat S_z\hat S_+\hat S_z
-\hat S_+\hat S_+\hat S_-
\Big),
\label{eq:Splus_Lminus_exact}
\end{equation}
where we have used Eq. (\ref{eq:su2}) repeatedly to reorder products.

Putting Eqs.~\eqref{eq:Splus_H}--\eqref{eq:Splus_Lminus_exact} together gives
\begin{equation}
\dot{\hat S}_+
= i\Delta \hat S_+ +\frac{i\epsilon}{2}\hat S_z
-\frac{\Gamma_+}{S}\hat S_z\hat S_+
+\frac{\Gamma_-}{2S^3}\Big(
2\hat S_z\hat S_+\hat S_z^2
+2\hat S_z\hat S_+\hat S_z
-\hat S_+\hat S_+\hat S_-
\Big).
\label{eq:Splus_full}
\end{equation}

Similarly, for $\hat X=\hat S_z$ one finds
\begin{equation}
\dot{\hat S}_z
=-\frac{\epsilon}{2}\hat S_y
+\frac{\Gamma_+}{S}\hat S_-\hat S_+
-\frac{\Gamma_-}{S^3}\hat S_+\hat S_-\hat S_z^2 ,
\label{eq:Sz_full}
\end{equation}
where the Hamiltonian part gives $i[\hat H,\hat S_z]=-(\epsilon/2)\hat S_y$ and the two dissipators follow from Eq.~\eqref{eq:heis_lindblad}.

\paragraph{Step 2: mean-field closure and large-$S$ scaling.}
We introduce the normalized magnetization components
\begin{equation}
m_\alpha \equiv \frac{\langle \hat S_\alpha\rangle}{S}\qquad (\alpha=x,y,z),\qquad
m_\pm\equiv \frac{\langle \hat S_\pm\rangle}{S}=m_x\pm i m_y .
\label{eq:mdef}
\end{equation}
In the thermodynamic (large-$S$) limit we apply a standard mean-field factorization for products of collective operators,
\begin{equation}
\langle \hat A \hat B\rangle \approx \langle \hat A\rangle\langle \hat B\rangle,\qquad
\langle \hat A \hat B \hat C\rangle \approx \langle \hat A\rangle\langle \hat B\rangle\langle \hat C\rangle,
\label{eq:mf_fact}
\end{equation}
and, crucially, we keep only the leading contributions in $S$.
This matters for the nonlinear-loss term in Eq.~\eqref{eq:Splus_full}:
\begin{itemize}
\item $\langle \hat S_z\hat S_+\hat S_z^2\rangle \sim \mathcal O(S^4)$, hence it survives after the prefactor $\Gamma_-/S^3$ and the final division by $S$.
\item $\langle \hat S_z\hat S_+\hat S_z\rangle \sim \mathcal O(S^3)$ and
$\langle \hat S_+\hat S_+\hat S_-\rangle\sim \mathcal O(S^3)$, hence they contribute only $\mathcal O(1/S)$ to $\dot m_+$ and are negligible as $S\to\infty$.
\end{itemize}
Therefore, at leading order,
\begin{align}
\frac{d m_+}{dt}
&=\frac{1}{S}\frac{d}{dt}\langle \hat S_+\rangle
\simeq
i\Delta m_+ +\frac{i\epsilon}{2}m_z
-\Gamma_+ m_z m_+
+\Gamma_- m_z^3 m_+ .
\label{eq:mp_mf}
\end{align}
In the same spirit, from Eq.~\eqref{eq:Sz_full} we use $\langle \hat S_-\hat S_+\rangle/S^2\simeq m_-m_+=m_x^2+m_y^2$ and
$\langle \hat S_+\hat S_-\hat S_z^2\rangle/S^4\simeq (m_x^2+m_y^2)m_z^2$ to obtain
\begin{equation}
\frac{d m_z}{dt}\simeq -\frac{\epsilon}{2}m_y
+\Gamma_+(m_x^2+m_y^2)-\Gamma_-(m_x^2+m_y^2)m_z^2 .
\label{eq:mz_mf_intermediate}
\end{equation}
Finally, using the Bloch-sphere constraint (proven below) $m_x^2+m_y^2=1-m_z^2$ in the large-$S$ mean-field dynamics,
Eq.~\eqref{eq:mz_mf_intermediate} becomes
\begin{equation}
\frac{d m_z}{dt}= -\frac{\epsilon}{2}m_y +(\Gamma_+-\Gamma_- m_z^2)(1-m_z^2).
\label{eq:mz_mf}
\end{equation}

\paragraph{Step 3: equations for $m_x,m_y,m_z$.}
Taking real and imaginary parts of Eq.~\eqref{eq:mp_mf} (recall $m_+=m_x+i m_y$) gives the closed nonlinear mean-field equations
\begin{align}
\dot m_x &= -\Delta m_y-(\Gamma_+-\Gamma_- m_z^2)m_x m_z ,
\label{eq:mx_mf}\\
\dot m_y &= \Delta m_x+\frac{\epsilon}{2}m_z-(\Gamma_+-\Gamma_- m_z^2)m_y m_z ,
\label{eq:my_mf}\\
\dot m_z &= -\frac{\epsilon}{2}m_y+(\Gamma_+-\Gamma_- m_z^2)(1-m_z^2).
\label{eq:mz_mf_final}
\end{align}
These are precisely Eqs.~(2) in the main text (with $\Gamma_\pm\equiv \Gamma_\pm$ there, if one uses $\Gamma$ notation).

\subsection{Dynamical equation for the spin-length}\label{sec:2_norm}
The mean-field flow preserves the Bloch-sphere constraint.
Defining $\vec m=(m_x,m_y,m_z)$, we compute
\begin{equation}
\frac{d}{dt}|\vec m|^2
=\frac{d}{dt}(m_x^2+m_y^2+m_z^2)
=2m_x\dot m_x+2m_y\dot m_y+2m_z\dot m_z .
\end{equation}
Substituting Eqs.~\eqref{eq:mx_mf}--\eqref{eq:mz_mf_final} and simplifying, the Hamiltonian terms cancel identically, and we obtain
\begin{equation}
\frac{d}{dt}|\vec m|^2
=2m_z\big(|\vec m|^2-1\big)\big(\Gamma_+-\Gamma_- m_z^2\big).
\label{eq:norm_flow}
\end{equation}
Hence, if $|\vec m(0)|^2=1$, then Eq.~\eqref{eq:norm_flow} implies $|\vec m(t)|^2=1$ for all times, which provides a convenient numerical consistency check when integrating the mean-field equations.

\section{Resonant threshold from mean-field linear stability}\label{sec:stab_eps_c}

Here we derive the critical driving strength $\epsilon_c$ under resonant driving,
$\Delta=\omega_0-\omega_d=0$, for the mean-field equations (main text, Eq.~(2)):
\begin{align}
\dot m_x &= -(\Gamma_+-\Gamma_- m_z^2)m_x m_z, \label{eq:mf_mx_res}\\
\dot m_y &= \frac{\epsilon}{2}m_z-(\Gamma_+-\Gamma_- m_z^2)m_y m_z, \label{eq:mf_my_res}\\
\dot m_z &= -\frac{\epsilon}{2}m_y+(\Gamma_+-\Gamma_- m_z^2)(1-m_z^2). \label{eq:mf_mz_res}
\end{align}
For $\Delta=0$, the manifold $m_x=0$ is invariant since $\dot m_x\propto m_x$.
Therefore the stability/threshold can be obtained from the reduced $(m_y,m_z)$ dynamics.

\paragraph{Stationary (QS) solution.}
Let $(m_y^\ast,m_z^\ast)$ denote a stationary point with $m_z^\ast\neq 0$.
From $\dot m_y=0$ we obtain
\begin{equation}
0=\frac{\epsilon}{2}m_z^\ast-(\Gamma_+-\Gamma_- (m_z^\ast)^2)m_y^\ast m_z^\ast
\quad\Rightarrow\quad
m_y^\ast=\frac{\epsilon}{2\left(\Gamma_+-\Gamma_- (m_z^\ast)^2\right)}.
\label{eq:my_star}
\end{equation}
Substituting Eq.~\eqref{eq:my_star} into $\dot m_z=0$ yields a single self-consistency condition for $m_z^\ast$:
\begin{equation}
\left(\Gamma_+-\Gamma_- (m_z^\ast)^2\right)^2\left(1-(m_z^\ast)^2\right)=\frac{\epsilon^2}{4}.
\label{eq:mz_star_condition_general}
\end{equation}
Thus, a real stationary (QS) solution exists only if the right-hand side does not exceed the maximum of the left-hand side over $(m_z^\ast)^2\in[0,1]$.

\paragraph{Resonant critical drive for Fig.~3 parameters.}
For Fig.~3 we use $\Gamma_+=1$ and $\Gamma_-=2$ (and $\Delta=0$).
Writing $x\equiv (m_z^\ast)^2\in[0,1]$, Eq.~\eqref{eq:mz_star_condition_general} becomes
\begin{equation}
(1-2x)^2(1-x)=\frac{\epsilon^2}{4}.
\label{eq:g_of_x}
\end{equation}
Define $g(x)=(1-2x)^2(1-x)$. Its derivative is
\begin{equation}
g'(x)=-(2x-1)(6x-5),
\end{equation}
so the only interior extrema are at $x=1/2$ and $x=5/6$, where $g(1/2)=0$ and $g(5/6)=2/27$.
On the boundaries, $g(0)=1$ and $g(1)=0$.
Therefore the global maximum is $g_{\max}=1$ attained at $x=0$.
Equation~\eqref{eq:g_of_x} admits a real stationary solution only if
\begin{equation}
\frac{\epsilon^2}{4}\le g_{\max}=1
\quad\Rightarrow\quad
\epsilon\le \epsilon_c=2.
\label{eq:eps_c_resonant}
\end{equation}
Hence, for $\Gamma_+=1$ the mean-field stability threshold is
\begin{equation}
\epsilon_c=2.0,
\end{equation}
consistent with the linear-stability analysis reported in Ref.~[54] and with the finite-size scaling extracted in Fig.~3.

\paragraph{Interpretation.}
For $\epsilon<\epsilon_c$, Eqs.~\eqref{eq:mf_my_res}--\eqref{eq:mf_mz_res} admit a stationary solution
$(m_y^\ast,m_z^\ast)$ satisfying Eq.~\eqref{eq:mz_star_condition_general};
this fixed point is the mean-field attractor corresponding to the QS phase.
When $\epsilon>\epsilon_c$, the stationary solution ceases to exist, and the dynamics is necessarily attracted to a time-periodic mean-field limit cycle, corresponding to the BTC phase in the thermodynamic limit.

\section{Physical picture from the angular equations: pendulum versus gyroscope}
\label{sec:physical_picture}

\subsection{Derivation of the angular equations on the Bloch sphere}
\label{subsec:derive_angles}

The mean-field dynamics in the rotating frame is conveniently expressed in terms of the normalized magnetization
$\bm m=(m_x,m_y,m_z)$ on the Bloch sphere, $|\bm m|^2=\sum_{\alpha=x,y,z}m_\alpha^2=1$.
Starting from Eq.~(2) in the main text,
\begin{align}
\dot m_x &= -\Delta m_y - \bigl(\Gamma_+ - \Gamma_- m_z^2\bigr)m_x m_z, \nonumber\\
\dot m_y &= \Delta m_x + \frac{\epsilon}{2} m_z - \bigl(\Gamma_+ - \Gamma_- m_z^2\bigr)m_y m_z, \label{eq:mxmy_mz_eom_pp}\\
\dot m_z &= -\frac{\epsilon}{2} m_y + \bigl(\Gamma_+ - \Gamma_- m_z^2\bigr)\bigl(1-m_z^2\bigr), \nonumber
\end{align}
we introduce spherical coordinates on the Bloch sphere,
\begin{equation}
m_x=\sin\theta\cos\phi,\qquad
m_y=\sin\theta\sin\phi,\qquad
m_z=\cos\theta,
\label{eq:spherical_param_pp}
\end{equation}
with $\theta\in[0,\pi]$ the polar (latitude) angle and $\phi\in[0,2\pi)$ the azimuthal phase.

\paragraph{Latitude equation.}
From $m_z=\cos\theta$ one has $\dot m_z=-\sin\theta\,\dot\theta$.
Using $1-m_z^2=\sin^2\theta$ and $m_y=\sin\theta\sin\phi$ in the $m_z$ equation yields
\begin{equation}
\dot\theta
=
\frac{\epsilon}{2}\sin\phi
-
\sin\theta\bigl(\Gamma_+-\Gamma_-\cos^2\theta\bigr),
\label{eq:theta_eom_pp}
\end{equation}
valid away from the poles ($\sin\theta\neq 0$).

\paragraph{Azimuth equation and exact cancellation of dissipation.}
A standard identity gives $m_x\dot m_y-m_y\dot m_x=\sin^2\theta\,\dot\phi$.
Applying Eqs.~(\ref{eq:mxmy_mz_eom_pp}) one finds
\begin{equation}
m_x\dot m_y-m_y\dot m_x
=\Delta(m_x^2+m_y^2)+\frac{\epsilon}{2}m_x m_z,
\end{equation}
where the dissipative terms cancel exactly.
With $m_x^2+m_y^2=\sin^2\theta$ and $m_x m_z=\sin\theta\cos\theta\cos\phi$, this yields
\begin{equation}
\dot\phi
=
\Delta+\frac{\epsilon}{2}\cot\theta\,\cos\phi,
\label{eq:phi_eom_pp}
\end{equation}
again for $\sin\theta\neq 0$.
Equations~(\ref{eq:theta_eom_pp}) and (\ref{eq:phi_eom_pp}) are the angular form discussed in the physics picture section in the main text.

\subsection{Meaning of each term and the unified physical picture}
\label{subsec:pp_terms}

Equations~(\ref{eq:theta_eom_pp}) and (\ref{eq:phi_eom_pp}) expose a transparent control logic by separating latitude (amplitude-like) and azimuth (phase-like) dynamics while keeping their feedback explicit.

\paragraph{Dissipation: nonlinear restoring force and the background attractor.}
The dissipative part of Eq.~(\ref{eq:theta_eom_pp}),
$-\sin\theta(\Gamma_+-\Gamma_-\cos^2\theta)$,
acts as a nonlinear latitudinal restoring force.
In the undriven limit ($\epsilon=0$), it generates either polar fixed points (PFPs) at the north/south poles or a finite-tilt self-sustained oscillator (SSO) background satisfying
$\cos\theta_0=\sqrt{\Gamma_+/\Gamma_-}$ for $\Gamma_+<\Gamma_-$.
In this sense, dissipation selects the \emph{background} on which the drive and detuning act.

\paragraph{Drive: phase-selecting torque with latitude amplification.}
The coherent drive enters Eq.~(\ref{eq:theta_eom_pp}) as a latitudinal thrust $(\epsilon/2)\sin\phi$,
and enters Eq.~(\ref{eq:phi_eom_pp}) as a phase-selecting torque $(\epsilon/2)\cot\theta\cos\phi$.
The prefactor $\cot\theta$ provides a strong \emph{latitude amplification}:
it diverges near a PFP ($\theta\to 0$ or $\pi$) but remains bounded on an SSO ($\theta\simeq\theta_0$).
This is the geometric reason why the same drive has qualitatively different effects on different backgrounds.

\paragraph{Resonant drive: from phase pinning to phase self-trapping and resonant BTC.}
On resonance ($\Delta=0$), the azimuthal motion is activated solely by the drive torque.
For weak driving, the phase-selecting term tends to favor $\cos\phi\simeq 0$, i.e., $\phi$ near $\pi/2$ (mod $\pi$),
which maximizes the thrust $(\epsilon/2)\sin\phi$ in Eq.~(\ref{eq:theta_eom_pp}).
This allows the thrust to balance the dissipative restoring force and stabilize a nontrivial dynamical state.

At stronger drive, this simple balance generically breaks down:
because the torque in Eq.~(\ref{eq:phi_eom_pp}) depends on $\cot\theta$,
the coupled $(\theta,\phi)$ dynamics generates an \emph{asymmetric effective phase potential} in which $\phi$ becomes \emph{self-trapped} (bounded oscillations around a biased phase).
The thrust $(\epsilon/2)\sin\phi(t)$ then becomes rapidly oscillating but biased, acquiring a finite time average.
This nonzero averaged thrust supports a stable limit cycle and yields a \emph{resonant BTC} for both backgrounds considered.

\paragraph{Detuning: constant precession torque and the fate of phase self-trapping.}
Detuning adds the constant drift $\Delta$ in Eq.~(\ref{eq:phi_eom_pp}), i.e., a uniform precession torque in the rotating frame.
Over long timescales this drift tends to make $\phi$ explore $[0,2\pi)$, thereby washing out $\phi$--$\theta$ correlations and suppressing the effective latitudinal thrust.
In this sense, detuning destroys phase self-trapping and causes the resonant BTC to \emph{melt} into a stationary attractor (a fixed point / phase-locked QS state).

\paragraph{Pendulum versus gyroscope: why the outcome depends on the background.}
On a PFP background, the divergent amplification $\cot\theta$ near the stable pole makes the response akin to a heavily damped pendulum:
the phase-selecting torque is extremely strong close to the pole, and the phase-locked fixed point remains robust over a wide detuning range.

By contrast, on an SSO background the amplification is bounded ($\cot\theta_0<\infty$) and the undriven phase is intrinsically free.
The same detuning term $\Delta$ can therefore destabilize the locked phase and induce sustained azimuthal motion.
Through the thrust term in Eq.~(\ref{eq:theta_eom_pp}), this azimuthal motion establishes a new self-consistent dynamical balance in $\theta$,
giving rise to a \emph{non-resonant BTC}.
This angular-equation picture is fully consistent with the structural distinction between PFP and SSO backgrounds emphasized in the main text.

\end{document}